\documentclass[pre,twocolumn,preprintnumbers,amsmath,amssymb]{revtex4}

\usepackage{graphicx}
\usepackage{dcolumn}
\usepackage{bm}

\begin{document}

\newcommand{\ie}{\textit{i.e. }}
\newcommand{\beqn}{\begin{eqnarray}}
\newcommand{\eeqn}{\end{eqnarray}}
\newcommand{\beq}{\begin{equation}}
\newcommand{\eeq}{\end{equation}}
\newcommand{\omegat}{\bar{\omega}}
\renewcommand{\bold}[1]{{\bf #1}}
\renewcommand{\r}{\bold{r}}
\newcommand{\tr}{{\tilde{r}}}
\renewcommand{\tt}{{\tilde{t}}}
\newcommand{\hatt}{{\hat{t}}}
\newcommand{\hatz}{{\hat{z}}}

\title{Periodic Pattern Formation in Evaporating Drops}

\author{Vladimir A. Belyi$^1$}
\email{vbelyi@polysci.umass.edu}
\author{D. Kaya$^2$}
\author{M. Muthukumar$^{1,2}$}
\affiliation{
$^1$Department of Polymer Science and Engineering, $^2$Department of Physics, University of Massachusetts, Amherst, MA 01003
}

\date{\today}

\begin{abstract}
Solute deposits from evaporating drops with pinned contact line are usually concentrated near the contact line. The stain, or pattern, left on the substrate then consists of a single ring, commonly known as a {\it coffee ring}. Here we report on a variation of this phenomenon when periodic patterns emerge. We attribute these to phase transitions in certain solutes as solute concentration increases. Examples may include dissolved to crystalline transition in salt or order-disorder transition in liquid crystals. Activated nature of the phase transitions, along with the newly imposed boundaries between phases, may then invert solute density profile and lead to periodic deposits. Hereby we develop a general theoretical model and report on experimental observations on salt in water.
\end{abstract}

\maketitle

Evaporation rate of sessile drops is generally limited by the rate of vapor diffusion in the surrounding atmosphere \cite{Deegan1997,Hu2005}. Faster evaporation near the contact line produces outward radial flow of liquid and solute. Residual solute then accumulates near the contact line and a ring of solute deposit is left on a substrate. This is known as the coffee-ring effect \cite{Deegan1997,Deegan2000,Govor2004}. Here we argue that situation changes once solute, such as salt or liquid crystal, allows for phase transitions. Once solute concentration exceeds the saturation density, solute suspension becomes unstable and a new phase is nucleated. Diffusion of solute near the phase boundary becomes relevant, along with the liquid induced solute flow, and concentric deposit rings may emerge.

\begin{figure}[tb]
\centerline{\includegraphics[width=2.4in,height=1.8in] {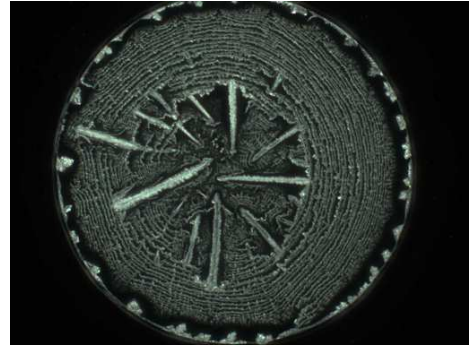}}
\caption{\label{fig::exppattern}Concentric rings observed in salt deposits of drying droplets. Drop diameter is approximately 2 mm. }
\end{figure}

Formation of solute rings may clearly be seen in salt deposits left by evaporating salt-water mixtures (Figure \ref{fig::exppattern}). For this experiment ternary solutions of sodium chloride (NaCl) and sodium polystyrenesulfonate (NaPSS) in deionized water were used \cite{Kaya2007}. Fixed-volume droplets were deposited onto glass substrate and allowed to evaporate. Salt deposits were then visualized with an optical microscope. We attribute the role of NaPSS in these experiments to pinning the droplet contact line, as discussed later in the text. 

\begin{figure*}[tb]
\begin{tabular}{ccc}
a) \includegraphics[width=2.1in,height=1.3in] {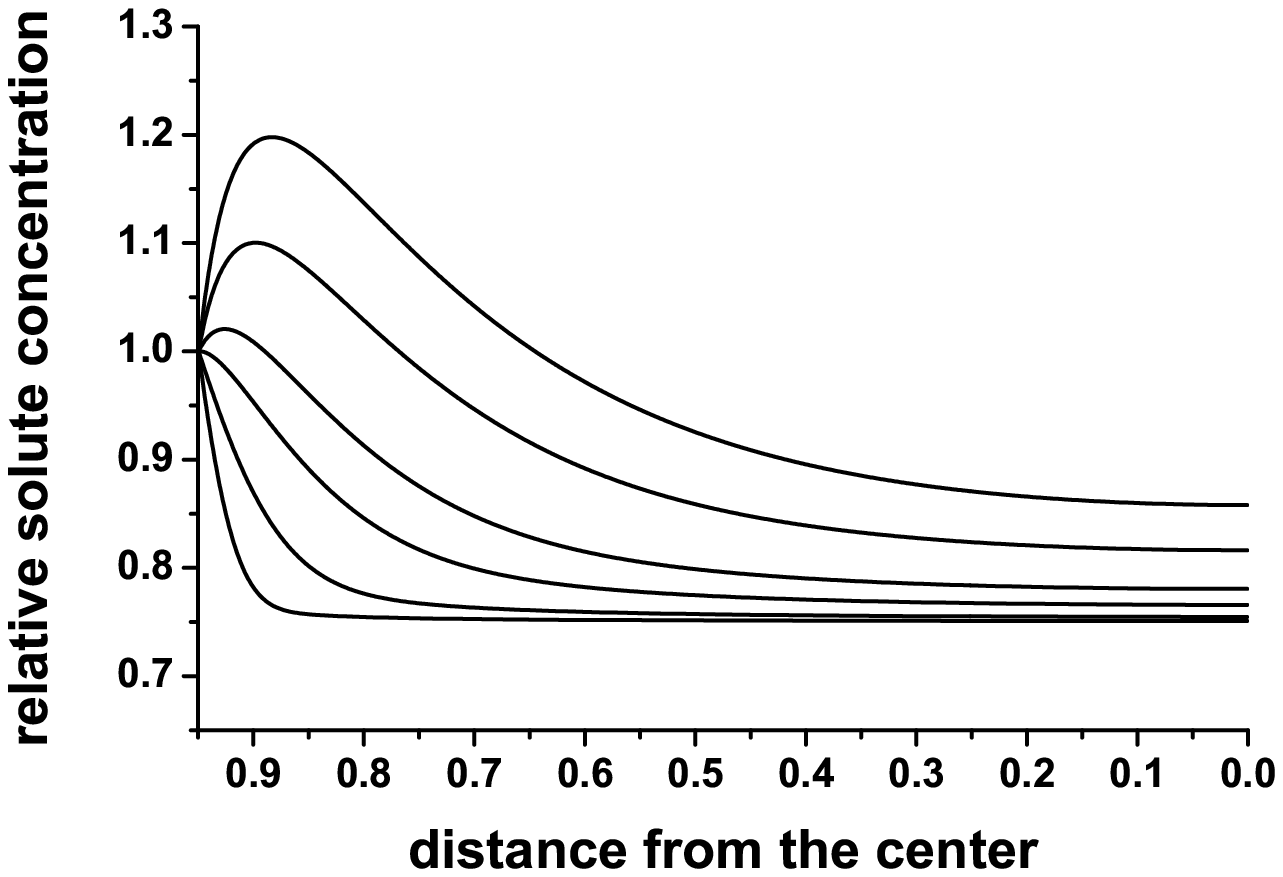} &
b) \includegraphics[width=2.1in,height=1.3in] {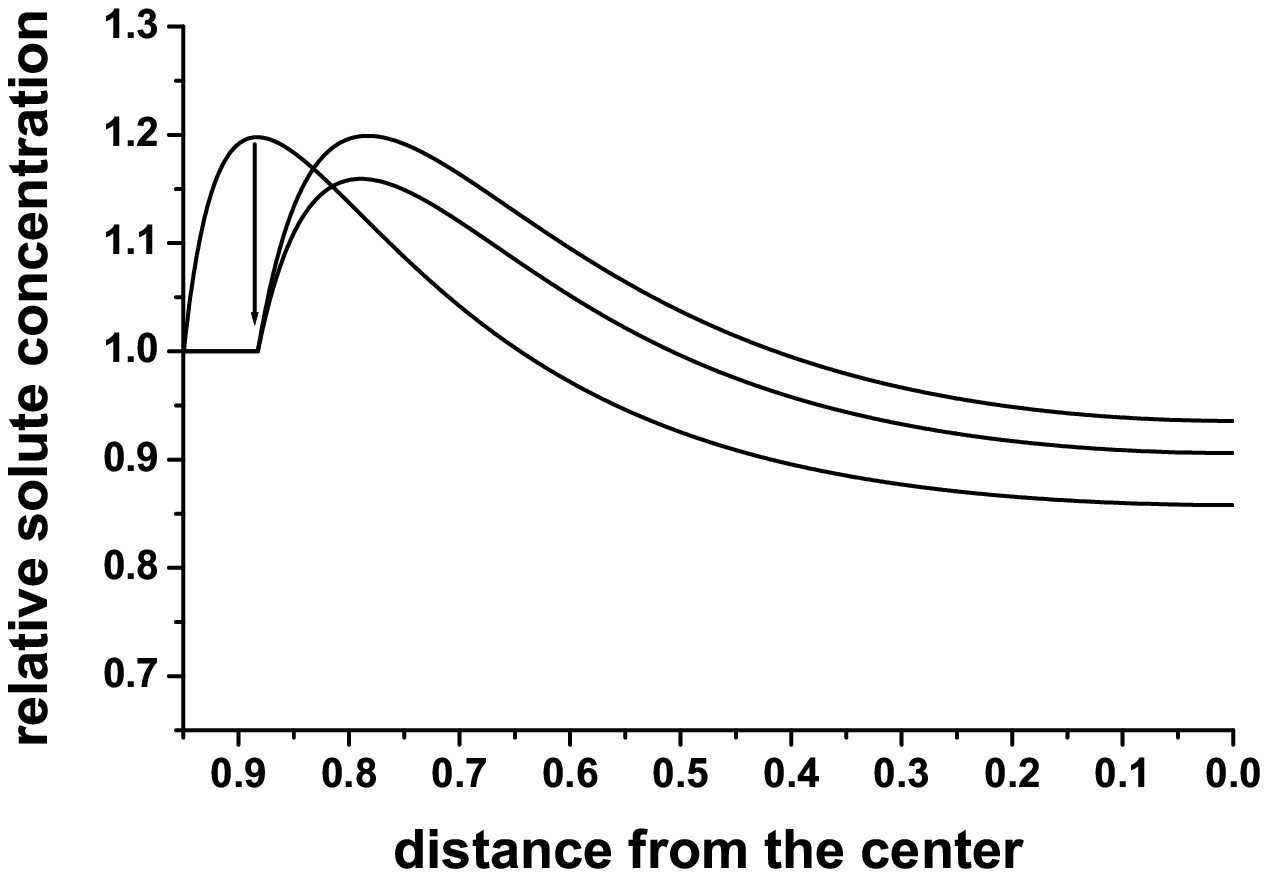} &
c) \includegraphics[width=2.1in,height=1.3in] {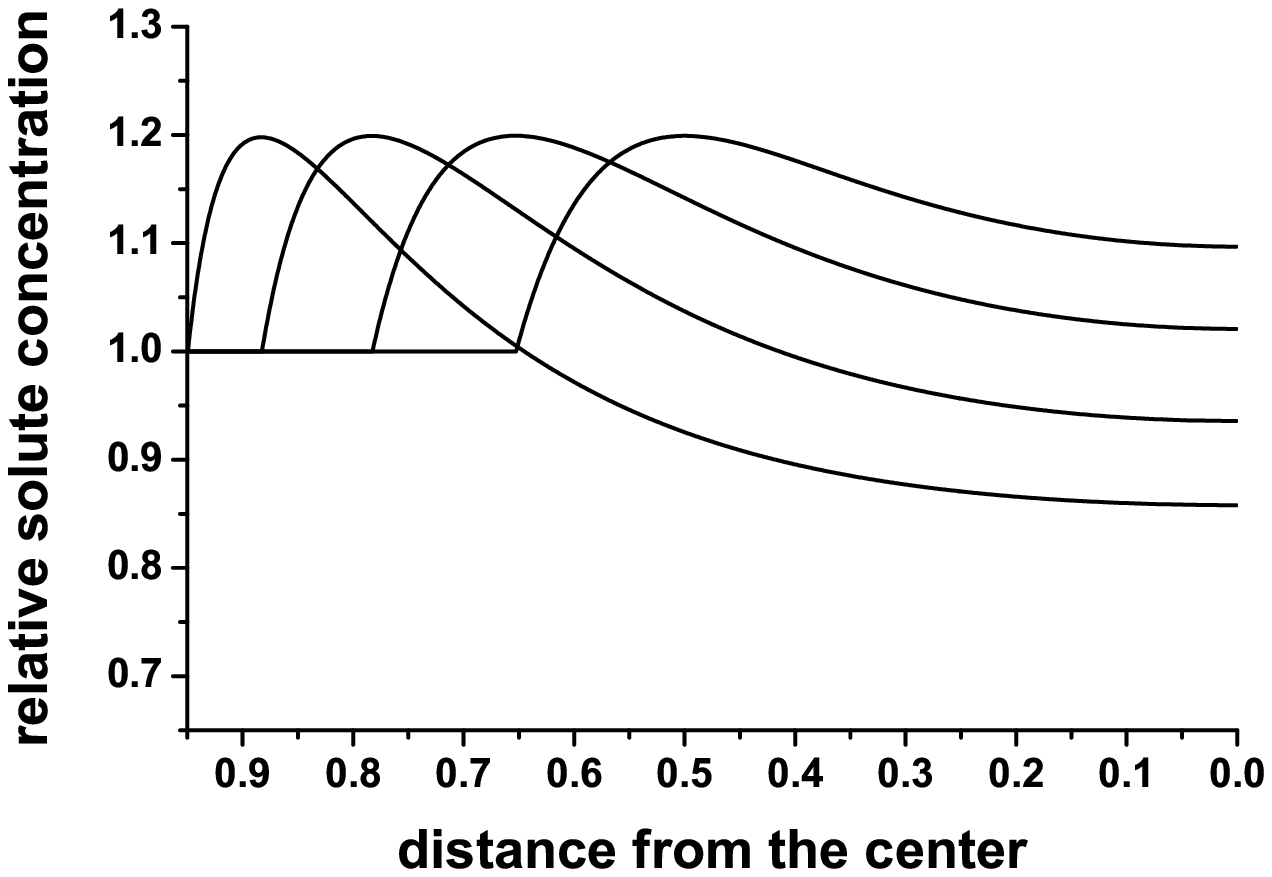}
\end{tabular}
\caption{\label{fig::ringformation}Schematic illustration of the concentric ring formation. Different curves represent solute density profiles at different times, with higher curves corresponding to progressively later time steps. Critical solute concentration for phase transitions is set at 1.0, and initial concentration is set at 0.75. Image a) illustrates increase in the solute concentration next to the first crystalline ring, assumed to be at radial coordinate $\tr = 0.95$. Solute concentration is fixed at critical value near the crystalline ring. Elsewhere in the drop concentration progressively increases with time producing oversaturated region; b) Once the peak concentration reaches suercritical value ($\phi_c = 1.20$ in this example), new crystalline ring is formed; at this moment phase boundary is moved to the location of the newly formed ring; c) Formation of progressive crystalline rings in the drop. }
\end{figure*}

Phase transitions render homogeneous solute suspension unstable above the critical concentration. Crystal nucleation or spinodal decomposition result in formation of new phases, such as salt crystals of Figure \ref{fig::exppattern}. In the evaporating drops, the first of these phase transitions takes place near the rim of the drop, in the region of the highest solute concentration. This is similar to the coffee-ring effect \cite{Deegan1997,Deegan2000,Govor2004}. Similarity, however, ends here as further flow of solute is affected by the boundary between phases. On the one hand, higher evaporation rate and evaporation induced flow of solute support higher solute concentration near the contact line. On the other hand, solute concentration at the phase boundary may not exceed the critical concentration. The solute concentration profile may then invert, with concentration decreasing near the contact line. This is schematically illustrated in Fig. \ref{fig::ringformation}a. Location of the peak in the solute concentration is the likely place for the subsequent ring nucleation. 

Proposed mechanism implies existence of the oversaturation region. Higher concentration of solute, often called {\it supercritical} concentration \cite{Henish1986c,HenishBook1970}, is needed for reliable phase nucleation. At concentrations just above critical, nucleation is possible, yet very slow. At the supercritical concentration, on the other hand, nucleation times become relevant relative to the evaporation times. Solute concentration may even reach the spinodal point so that phase transitions occur spontaneously. We will use the term nucleation irrespective of the actual origin of the supercritical concentration.

We now qualitatively estimate ring spacing $\Delta r$. Diffusive flow of solute near the contact line is $D \Delta \phi / \Delta r$, where $\Delta \phi$ is difference between critical and supercritical concentrations. Evaporation induced flow is $u_0 \phi \sim \phi R / T_0$, where $u_0 = R/T_0$ is characteristic velocity of the liquid, and $R$ and $T_0$ are radius of the drop and its evaporation time. Rings emerge when two flows become comparable, so that the typical ring spacing is $\Delta r \sim D T_0 \Delta \phi / R \phi$. This is of the order of tens of microns for salt in water. Concentric rings of this separation are readily resolved (Fig. \ref{fig::exppattern}).

Pattern formation described here complements other non-equilibrium phenomena observed in evaporating drops. Convective flows were speculated to modulate polymer deposits in evaporating drops \cite{Gonuguntla2004}. Surface forces and surface rupture were reported to order submicron sized particles \cite{Adachi1995,Ma2004,Mougin2002} and salt deposits \cite{Takhistov2002} in drying drops. Viscous drag was found to align DNA molecules near the contact line \cite{Smalyukh2006}. Yet none of these mechanisms deals with phase transitions in solutes. Proposed mechanism is also distinctly different from the Liesegang ring formation in gels \cite{Henish1986c,HenishBook1970}, which is also based on salt crystal nucleations. In the evaporating drops nucleation events do not require diffusion of two solute species, as in Liesegang rings. Instead, concentration gradients and nucleation events arise from the competition between solute diffusion and evaporation-driven liquid flow.

We now formulate our model. Comparison with experimental observations will follow. Sessile drop evaporation rate is primarily determined by the diffusion rate of liquid molecules in the atmosphere surrounding the drop, rather than by the rate at which these molecules escape the surface of the drop. This leads to the fundamental prediction that evaporation rate is highest next to the contact line, and decreases as one approaches the center of the drop \cite{Deegan1997}. In the limit of small contact angles and low solute concentrations the evaporation rate $J(r)$ is given by \cite{Popov2005}
\beq
J(r) = \frac{2}{\pi} \frac{D_{air} (\rho_{v} - \rho_{\infty})} {\sqrt{R^2  - r^2}}
\eeq
\noindent
where $D_{air}$ is diffusion constant for vapor particles in the atmosphere surrounding the drop, $\rho_v$ and $\rho_\infty$ are saturated and ambient vapor densities, and $R$ and $r$ are drop radius and distance from the center respectively. Throughout this paper we assume that concentration of solute is small enough not to affect liquid flow or evaporation rate. 

The total evaporation time for a drop with initial contact angle $\theta_0$ and liquid density $\rho$ is
\beq
T_0 = \frac{\text{drop mass}}{2 \pi \int_0^R J(r) r dr} = \frac{\pi \rho R^2}{16 D_{air} (\rho_{v} - \rho_{\infty})} \theta_0.
\eeq
This time is much longer than typical relaxation times inside the liquid. The drop maintains its spherical cap shape throughout the evaporation, with height profile $h(r) = \theta (R^2 - r^2) / 2R$. Here $\theta$ is contact angle, assumed small.

Radial velocity $u$ of the liquid in the drop may be readily evaluated from the constitutive equation 
\beq
\frac{1}{r} \nabla (r h u) + \frac{J}{\rho} + \partial_t h = 0,
\eeq
which immediately integrates to
\beq
u(r) = \frac{4 D_{air} (\rho_v - \rho_\infty)}{\pi \rho} \frac{1}{r \theta} \frac{\sqrt{1 - (r/R)^2} - [1 - (r/R)^2]^2}{1 - (r/R)^2},
\eeq

So far we have only considered flow of liquid. Solute particle, that generally move with the liquid, may also diffuse in it with concentration profile $\phi$ satisfying
\beq
\label{eq::phimain}
\partial_t (\phi h) + \frac{1}{r} \frac{\partial}{\partial r} \left[ r u h \phi - D h \frac{\partial \phi}{\partial r} \right] = 0,
\eeq
where $D$ is diffusion coefficient for solute particles. For the time being we assume that flow is one dimensional and ignore azimuthal coordinates. It is convenient to introduce dimensionless time $\tt = t / T_0$ and coordinate $\tr = r / R$. The main equation for the solute profile is therefore
\beq
\label{eq::main}
\begin{split}
\partial_t \left[ (1-\tt) \phi \right] = \\
- \frac{1}{\tr (1-\tr^2)} \partial_{\tr} \left\{ \frac{1}{4} \left[\sqrt{1-\tr^2} - (1-\tr^2)^2\right] \phi \right. \\
- \left. \Gamma \tr \left[ (1 - \tr^2) (1-\tt) \right] \partial_{\tr} \phi \right\}
\end{split}
\eeq
Typical value for the parameter $\Gamma = \frac{\pi}{16} \frac{D \rho}{D_{air} (\rho_{v} - \rho_{\infty}) } \theta_0$ at room temperature for salt diffusion in water is $\Gamma \sim 0.7 \theta_0$, assuming $\rho \approx 10^3$ kg/m$^3$, $\rho_v - \rho_\infty \approx 1.7 \cdot 10^{-2}$ kg/m$^3$, $D_{air} \approx 0.24 \cdot 10^{-4}$ m$^2$/s, $D \approx 1.5 \cdot 10^{-9}$ m$^2$/s.

At the onset of evaporation, solute flow follows the well-known coffee-ring phenomenon \cite{Deegan2000}. Diverging evaporation rate at the contact line creates excess in solute concentration, and the first ring nucleates near the contact line. We do not consider this effect here. Further flow of the solute proceeds with a new boundary condition $\phi|_{\tr=1-\delta} = 1$, which requires that solute concentration near the newly formed dense phase must equal the critical concentration. Here $\delta$ is location of the nucleated ring, and we have rescaled all solute concentrations in units of the critical concentration.

The most interesting aspect of the deposit is the spacing between rings. We start with numerical analysis, and calculations are carried out as follows. The initial density concentration $\phi_0$ is set at certain value between $\phi_0 < 1$, and adsorbing boundary condition is placed at $\tr = 1 - \delta$. The system is then allowed to evolve according to Eq. \ref{eq::main} until concentration reaches supercritical value $\phi_c$ at some radial coordinate $\tr$ (Fig. \ref{fig::ringformation}). Next crystalline ring is presumed to form at this position and adsorbing boundary is moved to the location of the newly formed ring. Calculations are then allowed to proceed. This process is schematically illustrated in Fig. \ref{fig::ringformation}. 

\begin{figure}[tb]
\begin{flushleft}
~~~~~a)\\
\end{flushleft}
\includegraphics[width=2.4in,height=1.5in] {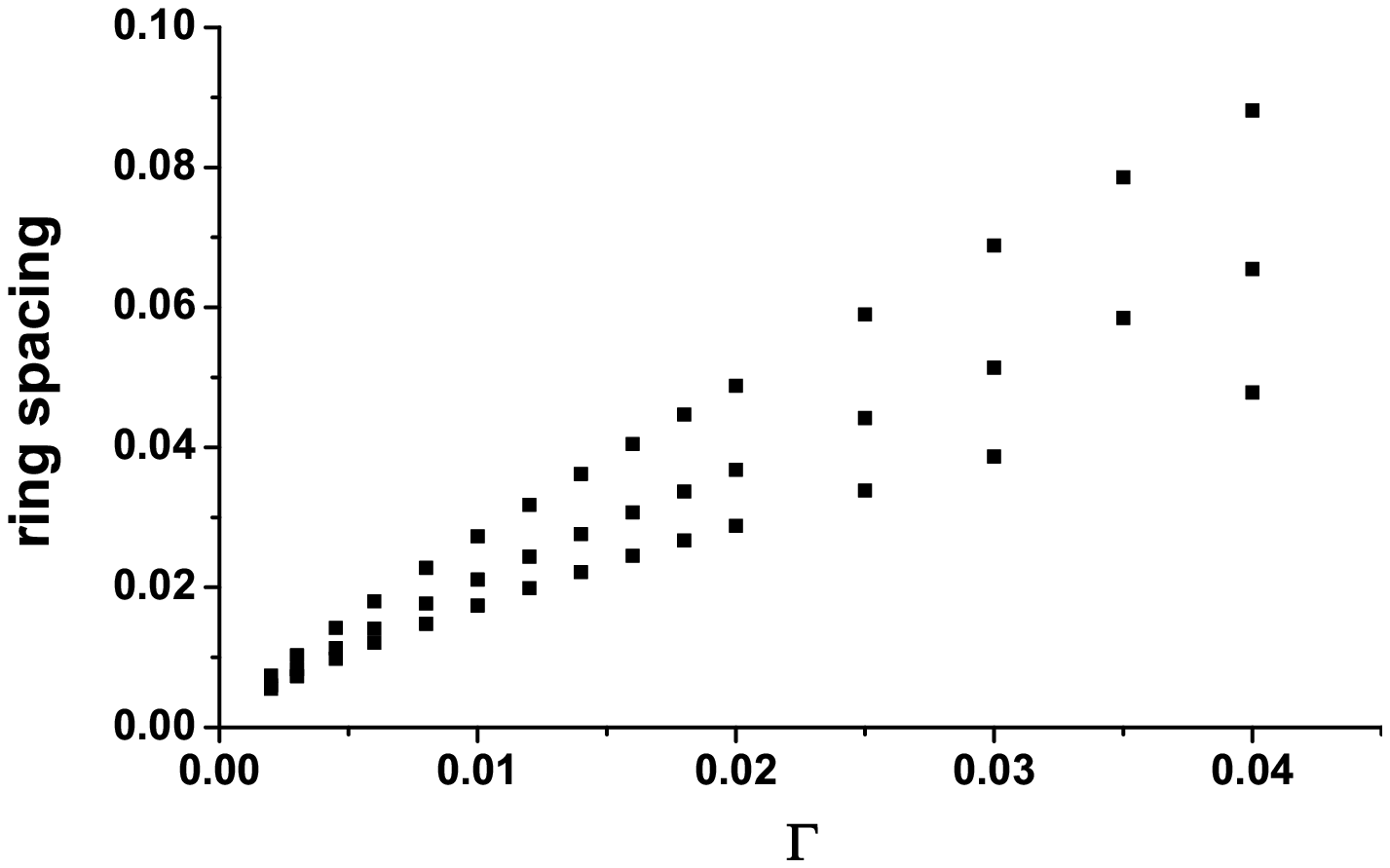} \\
\begin{flushleft}
~~~~~b)\\
\end{flushleft}
\includegraphics[width=2.4in,height=1.5in] {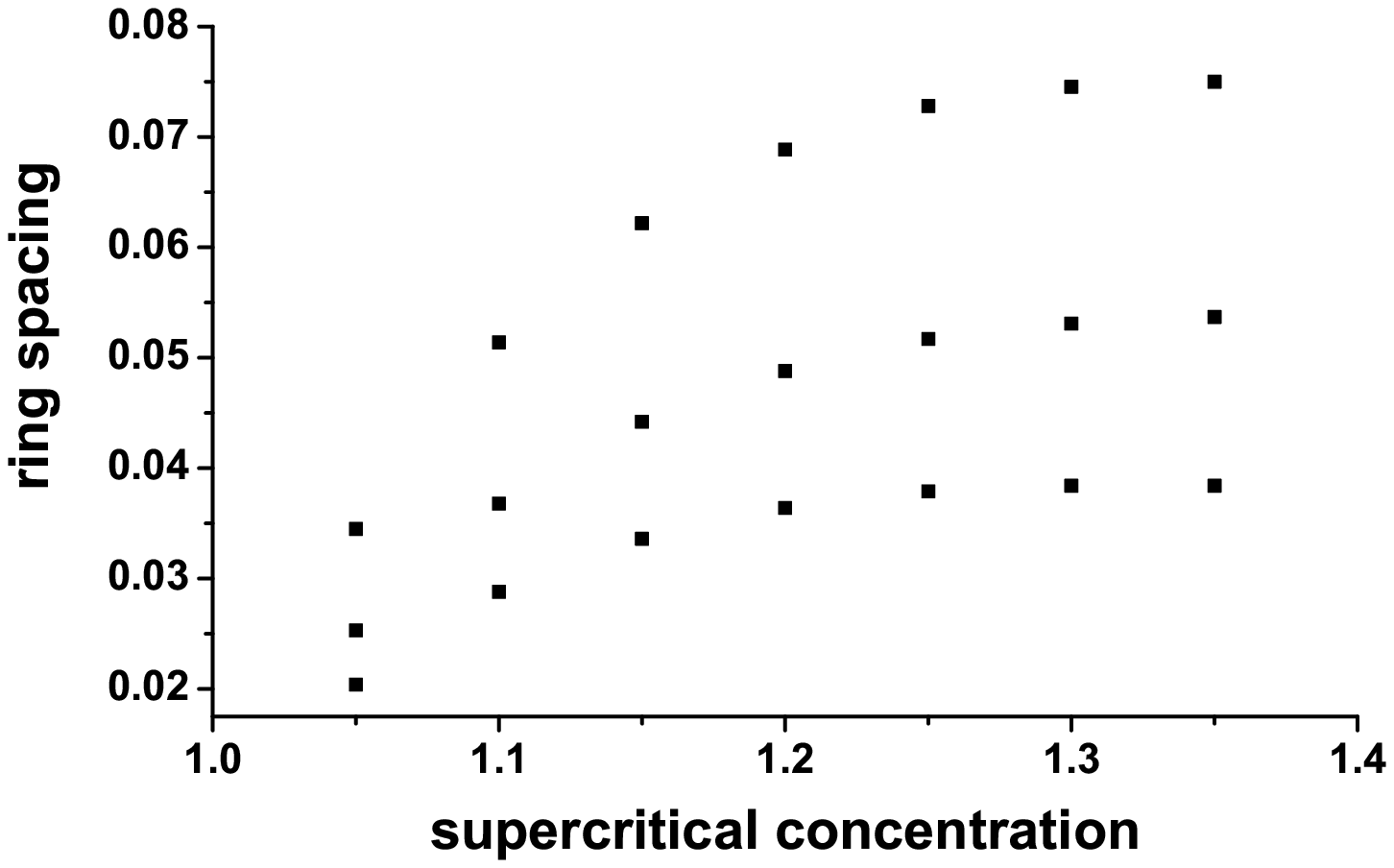}
\caption{\label{fig::spacings}Numerical calculations of ring spacing $\Delta \tr$. a) Change in ring spacing with $\Gamma$; curves, top to bottom, correspond to $\phi_c=1.2$ and $\phi_0=0.95$; $\phi_c=1.1$ and $\phi_0=0.95$; $\phi_c=1.1$ and $\phi_0=0.75$. b) Change in ring spacing with $\phi_c$. Curves, top to bottom, correspond to $\phi_0=0.95$ and $\Gamma = 0.03$; $\phi_0=0.95$ and $\Gamma = 0.02$; $\phi_0=0.75$ and $\Gamma = 0.02$.}
\end{figure}

In the numerical analysis, we varied parameters in the range $\Gamma = 0.002 - 0.05$, $\phi_0 = 0.05 - 0.95$, and $\phi_c = 1.05 - 1.40$. The grid size was set small enough to get stable solution, typically around $10^{-4}$ for coordinate step, and $10^{-8}$ for time step. The ring spacing $\Delta \tr$, which was found surprisingly uniform across large area of the drop surface, was measured as a function of these parameters, as shown in Fig. \ref{fig::spacings}. Particular attention was given to the range of spacings $\Delta \tr \sim 0.01 - 0.05$, which corresponds to experimentally achievable spacing for salt deposits discussed later in the text. Outside this range experimental system becomes unstable towards periodic pattern formation.

The first conclusion is that predicted ring spacing is roughly linear in $\Gamma$. This scaling can actually be extracted from the asymptotic expansion of Eq. \ref{eq::main}. We first introduce new variable $z = \sqrt{1 - r^2}$, which is closely related to the distance from contact line, and expand (\ref{eq::main}) in small $z$
\beq
\label{eq::mainsmallz}
\partial_\tt [ (1-\tt) \psi ] = \frac{1}{z^3} \partial_z \left[ \frac{1}{4} z \psi + \Gamma z (1 - \tt)  \partial_z \psi \right].
\eeq
This equation can be converted into parameter free form using substitution $\hatz = \Gamma z$ and $\hatt = \Gamma^3 \tt$. Leaving terms small in time and $\Gamma$,
\beq
\partial_\hatt \psi = \frac{1}{\hatz^3} \partial_\hatz \left[ \frac{1}{4} \hatz \psi + \hatz \partial_\hatz \psi \right].
\eeq
Spacing between rings at fixed location $\tr$ in the drop should therefore scale as $\Delta \tr \approx \Delta \hatz \sqrt{1 - \tr^2} / \tr \propto \Gamma$, in line with the numerical calculations (Fig. \ref{fig::spacings}a).

\begin{figure}[tb]
\includegraphics[width=2.4in,height=1.5in] {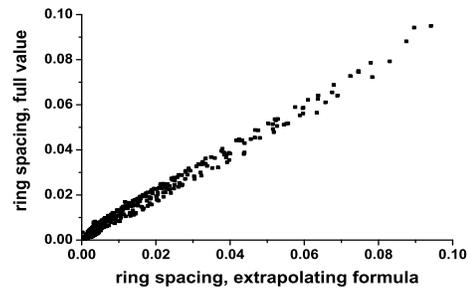}
\caption{\label{fig::numericfit}Comparison of numerically calculated ring spacing (vertical axis) versus predictions of interpolating expression (\ref{eq::interpolation}) (horizontal axis).}
\end{figure}

Concentration dependence of the ring spacing is richer. Experiments and numerical calculations show that ring nucleation events proceed in rapid succession. At that time average solute concentration is actually close to critical concentration (1.0 in our rescaled variables). That would imply that $1 - \tr \sim 1$ during early ring nucleation, and spacing should therefore scale as $\Delta \tr \propto \Gamma \phi_0$. Fitting to numerical data shows that this scaling is indeed close to the actual trend, with $\Delta \tr \sim A(\phi_0) \Gamma$, where $A(\phi_0) \approx \phi_0 - 0.036 - 13.97 (\phi_c - 1) \Gamma$.

Finally, we can add dependence on the supercritical concentration $\phi_c$. Assuming that ring spacing varies as
\beq
\label{eq::interpolation}
\Delta \tr = A(\phi_0) B(\phi_c - 1) \; \Gamma
\eeq
we find $B(\phi_c - 1) \approx 0.874 [1 + 14.42 (\phi_c - 1) - 19.10 (\phi_c - 1)^2]$. Comparison between this interpolating expression and full numerical calculation is given in Fig. \ref{fig::numericfit}.

Aforementioned predictions may now be compared with the experiments on salt deposit growth in evaporating salt-water mixtures. For this purpose we have used ternary solutions of sodium chloride (NaCl) and sodium polystyrenesulfonate (NaPSS) in deionized water. Fixed-volume droplets of this solution were deposited onto glass substrate and allowed to evaporate in a humidity-controlled glovebox. Glass substrates were {\it a priori} rinsed with RO water for a few minutes and dried in nitrogen atmosphere. Observed patterns (Fig. \ref{fig::exppattern}) where then analyzed for the ring spacing. We attribute the role of NaPSS in these experiments to pinning the droplet contact line. Neither degree of polymerization of NaPSS (varied in the range between 100,000 Da and 1,000,000 Da) nor its concentration (0.1 g/l to 1 g/l), seemed to affect the ring spacing. We refer the reader to Ref. \onlinecite{Kaya2007} for detailed discussion of the experimental procedures. 

\begin{figure}[tb]
\includegraphics[width=2.6in,height=1.7in] {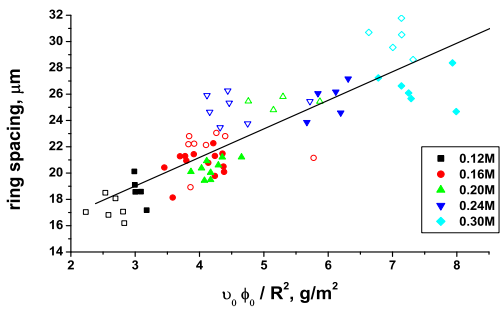}
\caption{\label{fig::expringspacing}Ring spacing in evaporated salt deposits. Open symbols correspond to NaPSS concentration 1 g/L, and filled symbols correspond to 0.1 g/L. The shape of each symbol is associated with the salt concentration, as shown in the legend. Uncertainty in measured values is estimated at 10\%. Solid line is a linear fit to the data points. }
\end{figure}

According to the predictions above, the spacing between salt rings is expected to follow $\Delta r \propto \Gamma \phi_0 R \propto \theta_0 \phi_0 R \propto v_0 \phi_0 / R^2$, where $v_0$, $\phi_0$, and $R$ are initial volume of the droplet, salt concentration, and radius of the droplet respectively. To test this prediction, we used drops with initial volume 0.5 $\mu$L and varied initial salt concentration between 120 and 300 mM. Additionally, glass substrates of varying cleanness were used to allow for larger spread in drop radii. Observed drops had radii in the range 1.0 - 1.3 mm. The results are summarized in Fig. \ref{fig::expringspacing}, and appear in good agreement with the above predictions. The uncertainty in the measured ring spacing and drop radius is estimated at 10\%. 

Observed trends in salt ring spacing may not be readily explained by other scenarios of ring formation. For example, nucleation of salt rings could cause surface rupture and contact line depinning \cite{Takhistov2002}. Dissolved polymer in the drop may then create repetitive pinning effect. In such a scenario, however, ring spacing would likely depend on polymer concentration rather than on salt concentration, contrary to the observations. Also, surface rupture is likely to favor unstable, fractal-like patterns \cite{Mougin2002,Reiter2001}. Another scenario could come from instabilities such as convective flow or Marangoni flow \cite{Gonuguntla2004}. Yet these flows are likely to be suppressed near the contact line, where polymer concentration is very large. These flows are also likely to be sensitive to polymer concentration, contrary to the observations \cite{Kaya2007}.

Finally, a few words on the stability of the concentric rings. In the discussion so far we have assumed radial symmetry of the problem. Yet nucleation of the crystalline rings occurs in the oversaturated regions of the drop. Growth of new phase should then be characterized by fingering instabilities of dendritic growth \cite{Langer1980,Warren1993,Grier1986}. In evaporating drops these instabilities are indeed possible, yet their radial extent is suppressed by the width of the oversaturated region (see Fig. \ref{fig::ringformation}b). Rapid decrease in solute concentration away from the supercritical point prevents instabilities from radial growth. 

Situation changes at late stages of evaporation, when the whole volume of the drop becomes oversaturated (Fig. \ref{fig::ringformation}c). Fingering instabilities are no longer constrained and may extend through the whole drop. This is indeed observed in salt deposits: in the inner region of the drop periodic ring formation is clearly replaced by radially oriented salt crystals or other disordered structures (Fig. \ref{fig::exppattern}). The transition between periodic rings and irregular structures in the inner region is common to all drops considered.

In conclusion, we have shown that phase transitions in solute may qualitatively change deposit growth in evaporating drops. The phase equilibrium between solute phases may invert the solute concentration profile, and lead to the growth of nearly periodic patterns. Formation of a single coffee-ring deposit is replaced by a sequence of concentric rings spanning large part of the drop area. The spacing between rings may be controlled via changes in ambient atmosphere, temperature, diffusion constant, and other parameters. This may prove crucial in understanding many contact line phenomena in evaporating liquids, and lead to better control over pattern growth.

This work was supported by NSF grant No 0605833 and the MRSEC at the University of Massachusetts Amherst.


\begin{thebibliography}{18}
\expandafter\ifx\csname natexlab\endcsname\relax\def\natexlab#1{#1}\fi
\expandafter\ifx\csname bibnamefont\endcsname\relax
  \def\bibnamefont#1{#1}\fi
\expandafter\ifx\csname bibfnamefont\endcsname\relax
  \def\bibfnamefont#1{#1}\fi
\expandafter\ifx\csname citenamefont\endcsname\relax
  \def\citenamefont#1{#1}\fi
\expandafter\ifx\csname url\endcsname\relax
  \def\url#1{\texttt{#1}}\fi
\expandafter\ifx\csname urlprefix\endcsname\relax\def\urlprefix{URL }\fi
\providecommand{\bibinfo}[2]{#2}
\providecommand{\eprint}[2][]{\url{#2}}

\bibitem[{\citenamefont{Deegan et~al.}(1997)\citenamefont{Deegan, Bakajin,
  Dupont, Huber, Nagel, and Witten}}]{Deegan1997}
\bibinfo{author}{\bibfnamefont{R.~D.} \bibnamefont{Deegan}},
  \bibinfo{author}{\bibfnamefont{O.}~\bibnamefont{Bakajin}},
  \bibinfo{author}{\bibfnamefont{T.~F.} \bibnamefont{Dupont}},
  \bibinfo{author}{\bibfnamefont{G.}~\bibnamefont{Huber}},
  \bibinfo{author}{\bibfnamefont{S.~R.} \bibnamefont{Nagel}}, \bibnamefont{and}
  \bibinfo{author}{\bibfnamefont{T.~A.} \bibnamefont{Witten}},
  \bibinfo{journal}{Nature} \textbf{\bibinfo{volume}{389}},
  \bibinfo{pages}{827} (\bibinfo{year}{1997}).

\bibitem[{\citenamefont{Hu and Larson}(2005)}]{Hu2005}
\bibinfo{author}{\bibfnamefont{H.}~\bibnamefont{Hu}} \bibnamefont{and}
  \bibinfo{author}{\bibfnamefont{R.~G.} \bibnamefont{Larson}},
  \bibinfo{journal}{Langmuir} \textbf{\bibinfo{volume}{21}},
  \bibinfo{pages}{3963} (\bibinfo{year}{2005}).

\bibitem[{\citenamefont{Deegan et~al.}(2000)\citenamefont{Deegan, Bakajin,
  Dupont, Huber, Nagel, and Witten}}]{Deegan2000}
\bibinfo{author}{\bibfnamefont{R.~D.} \bibnamefont{Deegan}},
  \bibinfo{author}{\bibfnamefont{O.}~\bibnamefont{Bakajin}},
  \bibinfo{author}{\bibfnamefont{T.~F.} \bibnamefont{Dupont}},
  \bibinfo{author}{\bibfnamefont{G.}~\bibnamefont{Huber}},
  \bibinfo{author}{\bibfnamefont{S.~R.} \bibnamefont{Nagel}}, \bibnamefont{and}
  \bibinfo{author}{\bibfnamefont{T.~A.} \bibnamefont{Witten}},
  \bibinfo{journal}{Phys. Rev. E} \textbf{\bibinfo{volume}{62}},
  \bibinfo{pages}{756} (\bibinfo{year}{2000}).

\bibitem[{\citenamefont{Govor et~al.}(2004)\citenamefont{Govor, Reiter, Parisi,
  and Bauer}}]{Govor2004}
\bibinfo{author}{\bibfnamefont{L.~V.} \bibnamefont{Govor}},
  \bibinfo{author}{\bibfnamefont{G.}~\bibnamefont{Reiter}},
  \bibinfo{author}{\bibfnamefont{J.}~\bibnamefont{Parisi}}, \bibnamefont{and}
  \bibinfo{author}{\bibfnamefont{G.~H.} \bibnamefont{Bauer}},
  \bibinfo{journal}{Phys. Rev. E} \textbf{\bibinfo{volume}{69}},
  \bibinfo{pages}{061609} (\bibinfo{year}{2004}).

\bibitem[{\citenamefont{Kaya et~al.}(2007)\citenamefont{Kaya, Belyi, and
  Muthukumar}}]{Kaya2007}
\bibinfo{author}{\bibfnamefont{D.}~\bibnamefont{Kaya}},
  \bibinfo{author}{\bibfnamefont{V.~A.} \bibnamefont{Belyi}}, \bibnamefont{and}
  \bibinfo{author}{\bibfnamefont{M.}~\bibnamefont{Muthukumar}},
  \bibinfo{journal}{unpublished}  (\bibinfo{year}{2007}).

\bibitem[{\citenamefont{Henish}(1986)}]{Henish1986c}
\bibinfo{author}{\bibfnamefont{H.~K.} \bibnamefont{Henish}},
  \bibinfo{journal}{J. Cryst. Growth} \textbf{\bibinfo{volume}{76}},
  \bibinfo{pages}{279} (\bibinfo{year}{1986}).

\bibitem[{\citenamefont{Henish}(1970)}]{HenishBook1970}
\bibinfo{author}{\bibfnamefont{H.~K.} \bibnamefont{Henish}},
  \emph{\bibinfo{title}{Crystal Growth in Gels}}
  (\bibinfo{publisher}{Pennsylvania State University Press, University Park,
  Pa}, \bibinfo{year}{1970}).

\bibitem[{\citenamefont{Gonuguntla and Sharma}(2004)}]{Gonuguntla2004}
\bibinfo{author}{\bibfnamefont{M.}~\bibnamefont{Gonuguntla}} \bibnamefont{and}
  \bibinfo{author}{\bibfnamefont{A.}~\bibnamefont{Sharma}},
  \bibinfo{journal}{Langmuir} \textbf{\bibinfo{volume}{20}},
  \bibinfo{pages}{3456} (\bibinfo{year}{2004}).

\bibitem[{\citenamefont{Adachi et~al.}(1995)\citenamefont{Adachi, Dimitrov, and
  Nagayama}}]{Adachi1995}
\bibinfo{author}{\bibfnamefont{E.}~\bibnamefont{Adachi}},
  \bibinfo{author}{\bibfnamefont{A.~S.} \bibnamefont{Dimitrov}},
  \bibnamefont{and} \bibinfo{author}{\bibfnamefont{K.}~\bibnamefont{Nagayama}},
  \bibinfo{journal}{Langmuir} \textbf{\bibinfo{volume}{11}},
  \bibinfo{pages}{1057} (\bibinfo{year}{1995}).

\bibitem[{\citenamefont{Ma et~al.}(2004)\citenamefont{Ma, Xia, Chen, Mi, Wang,
  and Shi}}]{Ma2004}
\bibinfo{author}{\bibfnamefont{X.}~\bibnamefont{Ma}},
  \bibinfo{author}{\bibfnamefont{Y.}~\bibnamefont{Xia}},
  \bibinfo{author}{\bibfnamefont{E.-Q.} \bibnamefont{Chen}},
  \bibinfo{author}{\bibfnamefont{Y.}~\bibnamefont{Mi}},
  \bibinfo{author}{\bibfnamefont{X.}~\bibnamefont{Wang}}, \bibnamefont{and}
  \bibinfo{author}{\bibfnamefont{A.-C.} \bibnamefont{Shi}},
  \bibinfo{journal}{Langmuir} \textbf{\bibinfo{volume}{20}},
  \bibinfo{pages}{9520} (\bibinfo{year}{2004}).

\bibitem[{\citenamefont{Mougin and Haidara}(2002)}]{Mougin2002}
\bibinfo{author}{\bibfnamefont{K.}~\bibnamefont{Mougin}} \bibnamefont{and}
  \bibinfo{author}{\bibfnamefont{H.}~\bibnamefont{Haidara}},
  \bibinfo{journal}{Langmuir} \textbf{\bibinfo{volume}{18}},
  \bibinfo{pages}{9566} (\bibinfo{year}{2002}).

\bibitem[{\citenamefont{Takhistov and Chang}(2002)}]{Takhistov2002}
\bibinfo{author}{\bibfnamefont{P.}~\bibnamefont{Takhistov}} \bibnamefont{and}
  \bibinfo{author}{\bibfnamefont{H.-C.} \bibnamefont{Chang}},
  \bibinfo{journal}{Ind. Eng. Chem. Res.} \textbf{\bibinfo{volume}{41}},
  \bibinfo{pages}{6256} (\bibinfo{year}{2002}).

\bibitem[{\citenamefont{Smalyukh et~al.}(2006)\citenamefont{Smalyukh, Zribi,
  Butler, Lavrentovich, and Wong}}]{Smalyukh2006}
\bibinfo{author}{\bibfnamefont{I.~I.} \bibnamefont{Smalyukh}},
  \bibinfo{author}{\bibfnamefont{O.~V.} \bibnamefont{Zribi}},
  \bibinfo{author}{\bibfnamefont{J.~C.} \bibnamefont{Butler}},
  \bibinfo{author}{\bibfnamefont{O.~D.} \bibnamefont{Lavrentovich}},
  \bibnamefont{and} \bibinfo{author}{\bibfnamefont{G.~C.~L.}
  \bibnamefont{Wong}}, \bibinfo{journal}{Phys. Rev. Lett.}
  \textbf{\bibinfo{volume}{96}}, \bibinfo{pages}{177801}
  (\bibinfo{year}{2006}).

\bibitem[{\citenamefont{Popov}(2005)}]{Popov2005}
\bibinfo{author}{\bibfnamefont{Y.~O.} \bibnamefont{Popov}},
  \bibinfo{journal}{Phys. Rev. E} \textbf{\bibinfo{volume}{71}},
  \bibinfo{pages}{036313} (\bibinfo{year}{2005}).

\bibitem[{\citenamefont{Reiter and Sharma}(2001)}]{Reiter2001}
\bibinfo{author}{\bibfnamefont{G.}~\bibnamefont{Reiter}} \bibnamefont{and}
  \bibinfo{author}{\bibfnamefont{A.}~\bibnamefont{Sharma}},
  \bibinfo{journal}{Phys. Rev. Lett.} \textbf{\bibinfo{volume}{87}},
  \bibinfo{pages}{166103} (\bibinfo{year}{2001}).

\bibitem[{\citenamefont{Langer}(1980)}]{Langer1980}
\bibinfo{author}{\bibfnamefont{J.~S.} \bibnamefont{Langer}},
  \bibinfo{journal}{Rev. Mod. Phys.} \textbf{\bibinfo{volume}{52}},
  \bibinfo{pages}{1} (\bibinfo{year}{1980}).

\bibitem[{\citenamefont{Warren and Langer}(1993)}]{Warren1993}
\bibinfo{author}{\bibfnamefont{J.~A.} \bibnamefont{Warren}} \bibnamefont{and}
  \bibinfo{author}{\bibfnamefont{J.~S.} \bibnamefont{Langer}},
  \bibinfo{journal}{Phys. Rev. E} \textbf{\bibinfo{volume}{47}},
  \bibinfo{pages}{2702} (\bibinfo{year}{1993}).

\bibitem[{\citenamefont{Grier et~al.}(1986)\citenamefont{Grier, Ben-Jacob,
  Clarke, and Sander}}]{Grier1986}
\bibinfo{author}{\bibfnamefont{D.}~\bibnamefont{Grier}},
  \bibinfo{author}{\bibfnamefont{E.}~\bibnamefont{Ben-Jacob}},
  \bibinfo{author}{\bibfnamefont{R.}~\bibnamefont{Clarke}}, \bibnamefont{and}
  \bibinfo{author}{\bibfnamefont{L.~M.} \bibnamefont{Sander}},
  \bibinfo{journal}{Phys. Rev. Lett.} \textbf{\bibinfo{volume}{56}},
  \bibinfo{pages}{1264} (\bibinfo{year}{1986}).

\end{thebibliography}
\end{document}